\documentclass[11pt]{article}
\usepackage{graphics}
\usepackage{axodraw}
\usepackage{psfrag}

\title{Stokes Vector of Photon in the Decays $B^0 \to \rho^0 \gamma$ and $B^0 \to K^{\ast} \gamma$}

\author{L. M. Sehgal and J. van Leusen \\ \it{Institute of Theoretical Physics, RWTH Aachen,} \\
\it{D-52056 Aachen, Germany}}

\date{}
\begin{document}

\maketitle

\begin{abstract}
We consider a model for the decay $\overline{B^0} \to \rho^0 \gamma$ in which the short-distance amplitude determined by
the Hamiltonian describing $b \to d \gamma$ is combined with a typical long-distance contribution $\overline{B^0} \to D^+ D^- \to \rho^0 \gamma$.
The latter possesses a significant dynamical phase which induces a $CP$-violating asymmetry $A_{\rm CP}$, as well as an
important modification of the Stokes vector of the photon. The components $S_1$ and $S_3$ of the Stokes vector 
$\vec{S} = (S_1, S_2, S_3)$ can be measured in the decay $\overline{B^0} \to \rho^0 \gamma^{\ast} \to \pi^+ \pi^- e^+ e^-$
where they produce a characteristic effect in the angular distribution $d \Gamma / d \phi$, $\phi$ being the angle between the
$\pi^+ \pi^-$ and $e^+ e^-$ planes. A similar analysis is carried out for the decays $\overline{B^0} \to \overline{K^{\ast}} \gamma$ and
$\overline{B^0} \to \overline{K^{\ast}} \gamma^{\ast} \to \pi^+ K^- e^+ e^-$.
\end{abstract}

\section{Introduction}

We study in this paper a long-distance contribution to the decay $\overline{B^0} \to \rho^0 \gamma$, which has the interesting
feature of possessing a large dynamical phase. When added to the short-distance amplitude determined by the $b \to d \gamma$
penguin operator, this produces an asymmetry $A_{\rm CP}$ between $\Gamma (\overline{B^0} \to \rho^0 \gamma)$ and 
$\Gamma(B^0 \to \rho^0 \gamma)$. In addition, the presence of the long-distance component affects the polarization state
(Stokes vector) of the photon. This effect can be measured in the decay $\overline{B^0} \to \rho^0 e^+ e^- \to \pi^+ \pi^- e^+ e^-$.
An analogous effect on the Stokes vector occurs in the decay $\overline{B^0} \to \overline{K^{\ast}} \gamma$. The Stokes vector turns
out to be very sensitive to the proposed long-distance contribution and thus may give more insight into the structure of the radiative 
decay amplitude.

The main contribution to the amplitude of the decay $\overline{B^0} \to \rho^0 \gamma$ is believed to come from the effective
Hamiltonian
\begin{equation}
H_{\rm eff} = \frac{G_F}{\sqrt{2}} V_{tb} V^{\ast}_{td} c_7 {\cal O}_7. \label{BRGHam}
\end{equation}
$G_F$ is Fermi's constant, $V_{ij}$ the CKM matix elements, $c_7$ the Wilson coefficient and ${\cal O}_7$ is the 
electromagnetic penguin operator
\begin{equation}
{\cal O}_7 = \frac{e}{8 \pi^2} \, \overline{d} \, \sigma_{\mu \nu} m_b (1 + \gamma_5) \, b \, F^{\mu \nu}.
\end{equation}
The corresponding amplitude contains a parity conserving (magnetic) term and a parity violating (electric) term
and can be written as \cite{Beyer01}:
\begin{equation}
{\cal A}(\overline{B^0} \to \rho^0 \gamma) = \frac{e G_F}{\sqrt{2}} \left( \epsilon_{\mu \nu \rho \sigma}
q_1^{\mu} \epsilon_1^{\ast \nu} q_2^{\rho} \epsilon_2^{\ast \sigma} M_{\rm SD}
+ i \epsilon_1^{\ast \mu} \epsilon_2^{\ast \nu} \left(
g_{\mu \nu} p \cdot q_1 - p_{\mu} q_{1 \nu} \right) E_{\rm SD} \right), \label{AmplitudePengDom}
\end{equation}
where $p$ is the momentum of the $\overline{B^0}$ meson, $q_1$ is the momentum of the photon and $\epsilon_1$ its
polarization vector, $q_2$ is the momentum of the $\rho^0$ meson and $\epsilon_2$ its polarization vector.
(The subscript $SD$ denotes short-distance.)

Using the identity $\sigma_{\mu \nu} = \frac{i}{2} \epsilon_{\mu \nu \alpha \beta} \sigma^{\alpha \beta} \gamma_5$ it
immediately follows that
\begin{equation}
E_{\rm SD} \equiv M_{\rm SD}. \label{ESDisMSD}
\end{equation}
Since only one weak phase is involved, this amplitude on its own produces no $CP$ violation. The branching ratio is:
\begin{equation}
Br(\overline{B^0} \to \rho^0 \gamma) = \frac{G_F^2 \alpha}{16} \, \tau_{B^0} \, m^3_{B^0} \left( 1 - \frac{m^2_{\rho^0}}{m^2_{B^0}}
\right) \left( \left| E_{\rm SD} \right|^2 + \left| M_{\rm SD} \right|^2 \right),
\end{equation}
where \cite{Beyer01}
\begin{equation}
M_{\rm SD} = - V_{tb} V^{\ast}_{td} c_7 \frac{m_b}{2 \pi^2} \frac{T^{B^- \to \rho^-}_1(0)}{\sqrt{2}}
\end{equation}
and $T^{B^- \to \rho^-}_1(0)$ is the form factor of the $B^- \to \rho^-$ transition due to the tensor current \cite{FFPapers}.

The decay $\overline{B^0} \to \rho^0 \gamma$ possesses another observable: the Stokes vector, specifiying the polarization
state of the photon. Conforming to the notation of Ref. \cite{Sehgal;Leusen}, we rewrite the decay amplitude (\ref{AmplitudePengDom})
in the general form
\begin{equation}
{\cal A}(\overline{B^0} \to \rho^0 \gamma) = \frac{e G_F}{\sqrt{2}} \left( \epsilon_{\mu \nu \rho \sigma}
q_1^{\mu} \epsilon_1^{\ast \nu} q_2^{\rho} \epsilon_2^{\ast \sigma} {\cal M}
+ \epsilon_1^{\ast \mu} \epsilon_2^{\ast \nu} \left(
g_{\mu \nu} p \cdot q_1 - p_{\mu} q_{1 \nu} \right) {\cal E} \right),
\end{equation}
where ${\cal M} = M_{\rm SD} + \ldots $ and ${\cal E} = i(E_{\rm SD} + \ldots )$, the dots
denoting further interaction terms to be introduced later. The polarization of the photon is defined by the density 
matrix $\rho$:
\begin{eqnarray}
\rho & = & \left( \begin{array}{cc}
|{\cal E}|^2 & {\cal E}^{\ast}{\cal M}\\
{\cal E}{\cal M}^{\ast} & |{\cal M}|^2
\end{array} \right) \label{StokesV} \\
& = & \frac{1}{2} \left( |{\cal E}|^2 + |{\cal M}|^2 \right) \left(
 {\mathbf 1} + \vec{S} \cdot \vec{\tau} \right), \nonumber
\end{eqnarray}
where $\vec{\tau}$ are the Pauli matrices and $\vec{S}$ is the Stokes vector. The components of the Stokes vector are according
to Eq. (\ref{StokesV}):
\begin{eqnarray}
S_1 & = & \frac{2 \, {\rm Re}({\cal E}^{\ast}{\cal M})}{|{\cal E}|^2 + |{\cal M}|^2}, \nonumber \\
S_2 & = & \frac{2 \, {\rm Im}({\cal E}^{\ast}{\cal M})}{|{\cal E}|^2 + |{\cal M}|^2}, \label{DefStokes} \\
S_3 & = & \frac{|{\cal E}|^2 - |{\cal M}|^2}{|{\cal E}|^2 + |{\cal M}|^2}. \nonumber
\end{eqnarray}
The component $S_2$ describes the circular polarisation of the photon which has been discussed by Grinstein and Pirjol \cite{Grinstein;Ali}.
More interesting from
our point of view are the components $S_1$ and $S_3$ which can be measured indirectly by studying the Dalitz pair process
$\overline{B^0} \to \rho^0 \gamma^{\ast} \to \rho^0 e^+ e^-$. If the short-distance amplitude is the only contribution
there is no $CP$ asymmetry ($A_{\rm CP} = 0$) and the photon is purely left-handed polarized, the Stokes vector reducing
to the trivial form ($S_{1,3} = 0$, $S_2 = -1$).

In the next Section, we introduce an extra contribution to the $\overline{B^0} \to \rho^0 \gamma$ amplitude that carries
not only a different weak phase but a non-trivial dynamical (strong) phase, thereby generating non-vanishing values for the
observables $A_{\rm CP}$, $S_1$ and $S_3$.

\section{A long-distance contribution to $\overline{B^0} \to \rho^0 \gamma$}

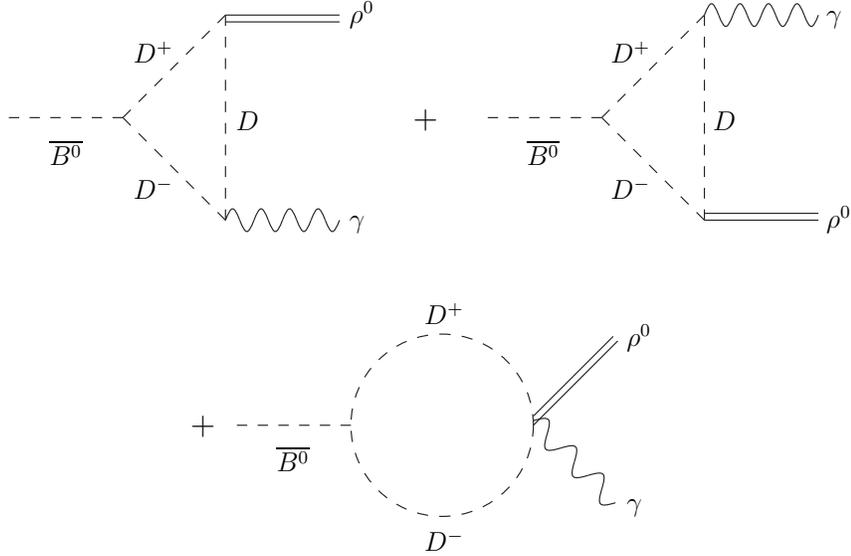
\begin{figure}
\center
\resizebox{12cm}{!}{
\begin{picture}(390,250)
\SetWidth{0.5}
\SetColor{Black}
\DashLine(0,190)(50,190){5}
\DashLine(50,190)(95,235){5}
\DashLine(95,235)(95,145){5}
\DashLine(95,145)(50,190){5}
\Photon(95,145)(145,145){5}{4}
\Line(95,235)(145,235)
\Line(95,232)(145,232)
\Text(18,170)[lb]{\large{$\overline{B^0}$}}
\Text(55,155)[lb]{\large{$D^-$}}
\Text(55,215)[lb]{\large{$D^+$}}
\Text(100,185)[lb]{\large{$D$}}
\Text(150,140)[lb]{\large{$\gamma$}}
\Text(150,230)[lb]{\large{$\rho^0$}}

\Text(178,185)[lb]{\Large{+}}

\DashLine(210,190)(260,190){5}
\DashLine(260,190)(305,235){5}
\DashLine(305,235)(305,145){5}
\DashLine(305,145)(260,190){5}
\Photon(305,235)(355,235){5}{4}
\Line(305,148)(355,148)
\Line(305,145)(355,145)
\Text(228,170)[lb]{\large{$\overline{B^0}$}}
\Text(265,155)[lb]{\large{$D^-$}}
\Text(265,215)[lb]{\large{$D^+$}}
\Text(310,185)[lb]{\large{$D$}}
\Text(360,230)[lb]{\large{$\gamma$}}
\Text(360,140)[lb]{\large{$\rho^0$}}

\Text(80,50)[lb]{\Large{+}}

\DashLine(100,55)(150,55){5}
\DashCArc(190,55)(40,0,360){5}
\Line(230,55)(267,92)
\Line(230,59)(265,94)
\Photon(230,57)(266,21){5}{3}
\Text(118,35)[lb]{\large{$\overline{B^0}$}}
\Text(272,15)[lb]{\large{$\gamma$}}
\Text(272,88)[lb]{\large{$\rho^0$}}
\Text(183,100)[lb]{\large{$D^+$}}
\Text(183,0)[lb]{\large{$D^-$}}
\end{picture}
}
\caption{Proposed long-distance contribution for $\overline{B^0} \to \rho^0 \gamma$: triangle, crossed and
sea-gull graph \label{LDtriangle}}
\end{figure}

We consider the long-distance contribution to the decay $\overline{B^0} \to \rho^0 \gamma$ depicted by the triangle graphs
in Fig. (\ref{LDtriangle}) with $D^+ D^-$ mesons as intermediate states\footnote{Similar graphs have been considered in connection with
the long-range contribution to $B^0 \to \gamma \gamma$; see, for example, Ref. \cite{CELZZ}. It may be mentioned that long-distance effects
of operators containing $\overline{c}c$ currents in charmless $B$-decays have also been discussed under the appelation {\it charming penguins}
\cite{CharmPeng}. We are not aware of a discussion of the radiative decays $\overline{B^0} \to \rho^0 \gamma$ 
($\overline{B^0} \to \overline{K^{\ast}}\gamma$) along these lines.}. The amplitude is calculated by analogy to the pion-loop model used for
the decay $K_S \to \gamma \gamma^{\ast}$, discussed in detail in Ref. \cite{CELZZ}. In such a model, based on minimal electromagnetic coupling, 
both real and imaginary parts of the amplitude are finite and calculable. Using dimensional regularization the gauge-invariant amplitude
for the three graphs (triangle + crossed + sea-gull) is purely electric:
\begin{equation}
{\cal A}_{\rm LD} = \frac{e G_F}{\sqrt{2}} i \epsilon_1^{\ast \mu} \epsilon_2^{\ast \nu} \left(
g_{\mu \nu} p \cdot q_1 - p_{\mu} q_{1 \nu} \right) E_{\rm LD}, \label{LDAmplitude}
\end{equation}
where
\begin{equation}
E_{\rm LD} = - V_{cb} V^{\ast}_{cd} \frac{2 f_{\rho^0 D^+ D^-} g_{B^0 D^+ D^-}}{(4 \pi )^2 m^2_{D^+}} \, F(m^2_{B^0}, m^2_{\rho^0})
\end{equation}
and \cite{FSmith}
\begin{eqnarray}
F(m^2_{B^0}, m^2_{\rho^0}) & = & - \frac{1}{2 (a - b)} +\frac{1}{(a - b)^2} \left[ \frac{f_a - f_b}{2} 
+ b \left( g_a - g_b \right) \right], \label{FB0Mrho} \\
f_a & = & - \left(\ln \left( \sqrt{a} + \sqrt{a-1} \right) - i \frac{\pi}{2} \right)^2, \nonumber \\
g_a & = & \sqrt{\frac{a-1}{a}} \left( \ln \left( \sqrt{a} + \sqrt{a-1} \right) - i \frac{\pi}{2} \right), \nonumber \\
f_b & = & \arcsin^2 \left( \sqrt{b} \right), \nonumber \\
g_b & = & \sqrt{\frac{1-b}{b}} \arcsin \left( \sqrt{b} \right), \nonumber \\
a & = & \frac{m^2_{B^0}}{4 m^2_{D^+}}, \nonumber \\
b & = & \frac{m^2_{\rho^0}}{4 m^2_{D^+}}. \nonumber
\end{eqnarray}
In this model there are only two parameters left: the coupling constants $g_{B^0 D^+ D^-}$ and $f_{\rho^0 D^+ D^-}$. 
The coupling $g_{B^0 D^+ D^-}$ is determined by data ($Br(\overline{B^0} \to D^+ D^-) = 2.46 \times 10^{-4}$) \cite{Browder}. 
For $f_{\rho^0 D^+ D^-}$ we use the vector dominance hypothesis, which implies $f_{\rho^0 D^+ D^-} = \frac{1}{2} f_{\rho^0 \pi^+ \pi^-}$. Using the
empirical value $f_{\rho \pi^+ \pi^-}^2 / 4 \pi \approx 2.5$, we thus have
\begin{eqnarray}
g_{B^0 D^+ D^-} & = & \frac{4}{G_F \left| V_{cb} V^{\ast}_{cd} \right|} 
\sqrt{\frac{2 \pi m_{B^0}}{\tau_{B^0} \sqrt{1- \frac{4 m^2_{D^+}}{m^2_{B^0}}}}} \sqrt{Br(\overline{B^0} \to D^+ D^-)},
\label{gDDlabel} \\
f_{\rho^0 D^+ D^-} & = & \frac{1}{2} f_{\rho \pi^+ \pi^-} \approx \sqrt{2.5 \pi}. \label{frholabel}
\end{eqnarray}

A comparison of the penguin amplitude with the above long-distance contribution reveals serveral interesting features.

(a) The two amplitudes have different CKM factors, hence different weak phases. In addition, the long-distance part has 
a large absorptive part, producing a significant strong phase:
\begin{equation}
\delta_{\rm dyn} = \arctan \left( \frac{{\rm Im} \left[ F(m^2_{B^0}, m^2_{\rho^0}) \right] }{{\rm Re} \left[ F(m^2_{B^0}, m^2_{\rho^0}) \right] } 
\right) \approx 97^{\circ}.
\end{equation}
This opens the way to a non-zero $CP$-violating asymmetry $A_{\rm CP}$.

(b) The long-distance component is quite sizeable in comparison to the short-distance amplitude. Taking the estimates in
Eqs. (\ref{frholabel}) and (\ref{gDDlabel}) at face value,
\begin{equation}
\frac{\Gamma_{\rm LD}}{\Gamma_{\rm SD}} \approx 30 \%.
\end{equation}

(c) The long-distance amplitude generated by the $D^+ D^-$ intermediate state is purely electric, in contrast to the equality
of $E_{\rm SD}$ and $M_{\rm SD}$ (Eq. (\ref{ESDisMSD})). This implies that the Stokes vector component $S_3$ will be non-zero.
The existence of the strong phase $\delta_{\rm dyn}$ also means that the component $S_1$ will be different from zero. Thus,
we can expect non-trivial effects associated with $S_{1,3} \neq 0$ in the Dalitz pair reaction 
$\overline{B^0} \to \rho^0 \gamma^{\ast} \to \rho^0 e^+ e^-$.

The amplitude ${\cal A}_{\rm LD}$ given by Eqs. (\ref{LDAmplitude}) - (\ref{FB0Mrho}), is based on minimal electromagnetic coupling,
and serves as a convenient reference value for the long-range contribution to $\overline{B^0} \to \rho^0 \gamma$, possessing
finite real and imaginary parts. The composite nature of the $D$-meson implies that there will be other intermediate states
such as $D D^{\ast}$, $D^{\ast} D^{\ast}$ etc., as well as possible form factor effects in the $D \overline{D}$ contribution.
Data on $B^0$ decays suggest that the $D D^{\ast}$, $D^{\ast} D^{\ast}$ final states are dominantly $CP = + 1$ \cite{BaBarColl},
the same as for $D^+ D^-$, implying that the effect of these intermediate states on $\overline{B^0} \to \rho^0 \gamma$
is mainly in the electric amplitude ${\cal E}$. In what follows we simulate the total long-distance amplitude by using
an expression of the form ${\cal A}_{\rm LD} = \xi {\cal A}_{\rm LD}(D^+ D^-)$, allowing the parameter $\xi$ to vary in the
range $-1 \leq \xi \leq 1$.

Inserting this parametrization in the Stokes vector in Eq. (\ref{DefStokes}) (neglecting the small $W$-exchange effects for clarity) the results are:
\begin{eqnarray}
S_1 & = & \frac{-2 \xi |E_{\rm SD}||E_{\rm LD}| \sin \left( \delta_{\rm dyn} - \beta \right) }{|{\cal E}|^2 + |{\cal M}|^2}, \nonumber \\
S_2 & = & \frac{-2 |E_{\rm SD} |^2 -2 \xi |E_{\rm SD}||E_{\rm LD}| \cos \left( \delta_{\rm dyn} - \beta \right) }{|{\cal E}|^2 + |{\cal M}|^2}, \\
S_3 & = & \frac{ \xi^2 |E_{\rm LD} |^2 + 2 \xi |E_{\rm SD}||E_{\rm LD}| \cos \left( \delta_{\rm dyn} - \beta \right) }{|{\cal E}|^2 + |{\cal M}|^2}. \nonumber
\end{eqnarray}
And the $CP$ asymmetry is ($A_{\rm CP} = (\Gamma(\overline{B^0}) - \Gamma(B^0))/(\Gamma(\overline{B^0}) + \Gamma(B^0))$)
\begin{equation}
A_{\rm CP} = \frac{4 \xi |E_{\rm SD}| |E_{\rm LD}| \sin \delta_{\rm dyn} \sin \beta}{|{\cal E}|^2 + |{\cal M}|^2 + |{\cal \overline{E}}|^2 + |{\cal \overline{M}}|^2}.
\end{equation}

\begin{figure}
\center
\psfrag{xi}[bl][bl][2]{$\xi$}
\resizebox{12cm}{!}{\rotatebox{-90}{\includegraphics{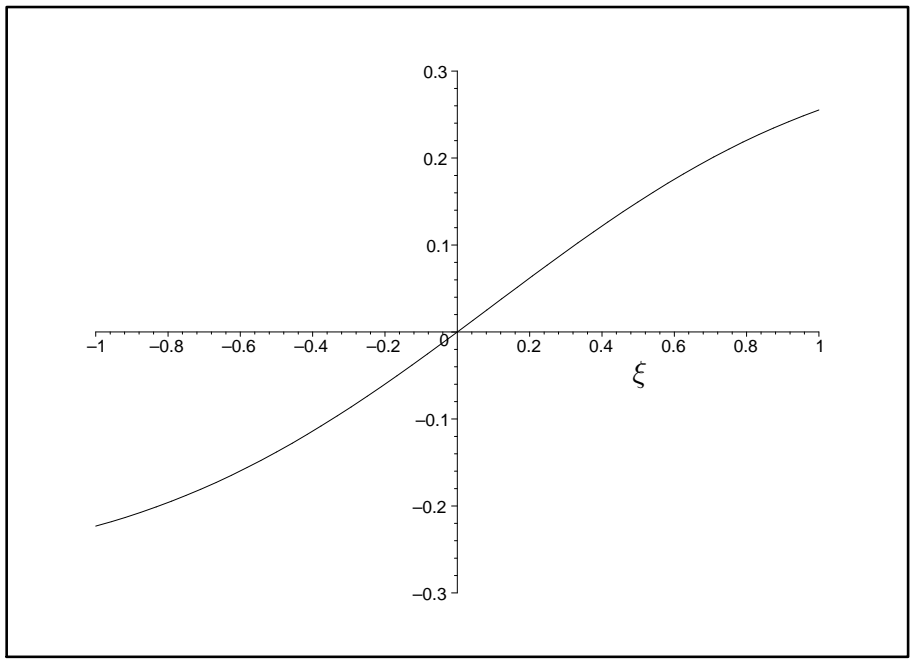}}}
\caption{The $CP$ asymmetry as a function of the scale parameter $\xi$ of the long distance contribution 
in $B \to \rho^0 \gamma$. \label{AsyPlot}}
\end{figure}

\begin{figure}
\center
\psfrag{xi}[bl][bl][2]{$\xi$}
\psfrag{a}[bl][bl][2]{(a)}
\psfrag{b}[bl][bl][2]{(b)}
\psfrag{c}[bl][bl][2]{(c)}
\resizebox{7cm}{!}{\rotatebox{-90}{\includegraphics{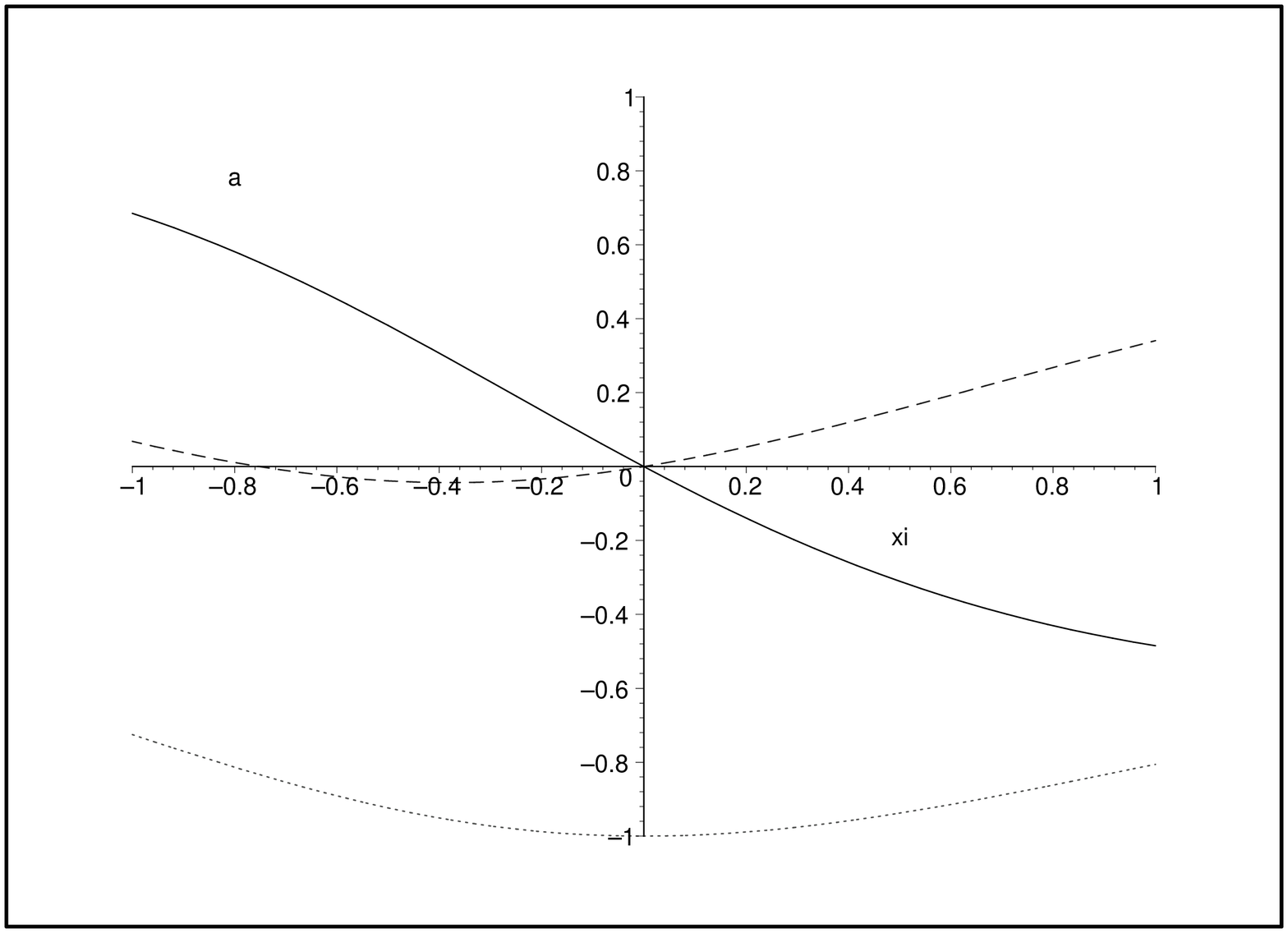}}}
\resizebox{7cm}{!}{\rotatebox{-90}{\includegraphics{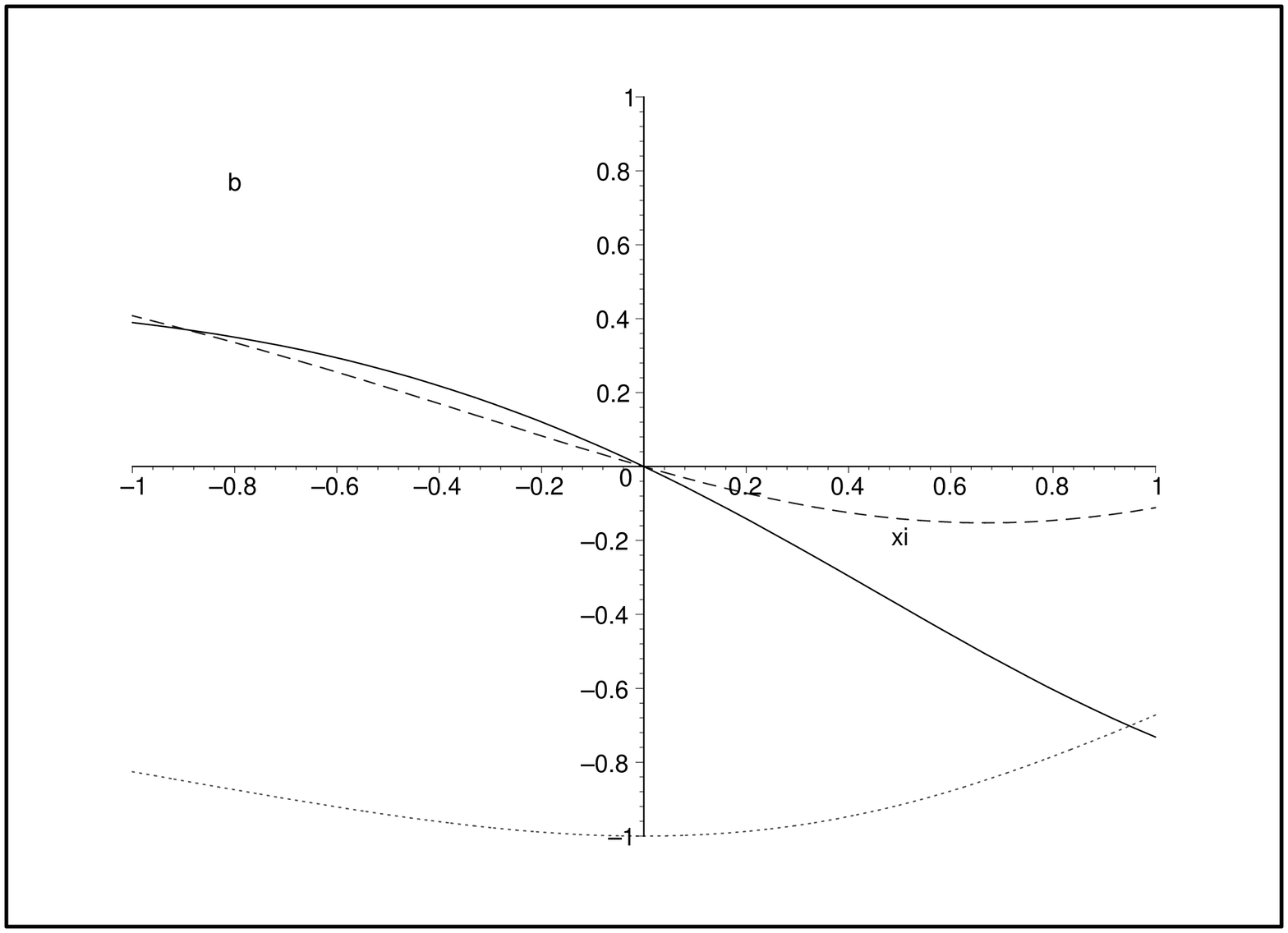}}}
\resizebox{7cm}{!}{\rotatebox{-90}{\includegraphics{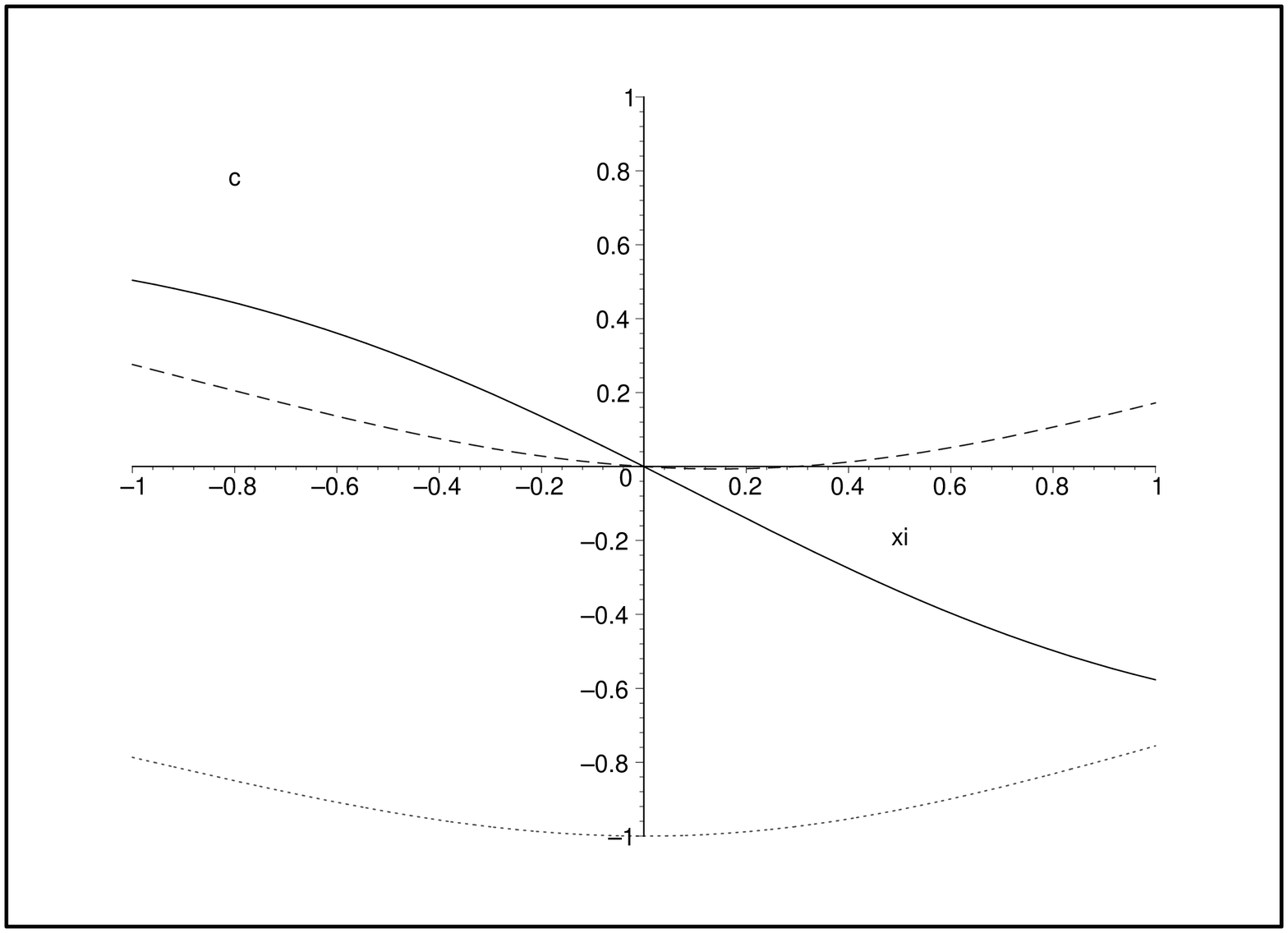}}}
\caption{The Stokes vector $\vec{S}$ as a function of the scale parameter $\xi$ of the long distance contribution 
in (a) $\overline{B^0} \to \rho^0 \gamma$, (b) $B^0 \to \rho^0 \gamma$ and (c) an untagged mixture of $\overline{B^0} / B^0 \to \rho^0 \gamma$. 
The solid line describes the $S_1$ component, the dotted line the $S_2$ component and the dashed line the $S_3$ component. \label{StokesPlotRG}}
\end{figure}

The results are shown in Fig. (\ref{AsyPlot}) for the $CP$ asymmetry and in Fig. (\ref{StokesPlotRG}) for the Stokes vector, respectively.
The central values used for the weak CKM phases are $\beta = 23^{\circ}$ and $\gamma = 59^{\circ}$.

The $CP$ asymmetry in Fig. (\ref{AsyPlot}) ranges from $-22 \%$ for
$\xi = -1$ to $26 \%$ for $\xi =1$ and vanishes at $\xi = 0$. Even for $|\xi| \approx 0.3$ which corresponds
to a long-distance contribution of $3 \%$ in the decay rate the asymmetry is still large: $|A_{\rm CP}| \approx 10 \%$. 

The effects of the long distance contribution to the Stokes vector are also large. The component $S_1$ (solid line) in 
Fig. (\ref{StokesPlotRG}) has a value around $0.7$ for $\xi = -1$ and around $-0.5$ for $\xi=1$ to be compared to $S_1 = 0$ at
$\xi = 0$. The component $S_3$ (dashed line) is small for $\xi < 0$ but approaches $0.35$ for $\xi = 1$.
The component $S_2$ (dotted line) is plotted for completeness.

The Stokes vector components $S_1$ and $S_3$ are observable in the decay 
$\overline{B^0} \to \rho^0 \gamma^{\ast} \to \pi^+ \pi^- e^+ e^-$ according to \cite{Sehgal;Leusen}
\begin{equation}
\frac{d \Gamma}{d s_l \, d \phi} \sim 1 - \left( \Sigma_1(s_l) \, \sin 2 \phi + \Sigma_3(s_l) \, \cos 2 \phi \right) , \label{PhiDist}
\end{equation}
where $\Sigma_1(0)$ and $\Sigma_3(0)$ are proportional to $S_1$ and $S_3$, respectively. $\phi$ is the angle between the dipion and 
the dilepton plane.

To determine the magnitude of $\Sigma_1$ and $\Sigma_3$ we follow reference \cite{KSSS}. There, the decay
$\overline{B^0} \to \pi^+ \pi^- e^+ e^-$ is constructed under the assumption that the pion pair is produced at the $\rho^0$
resonance in a narrow width approximation. Additional to the short-distance contribution due to ${\cal O}_7$ in the Hamiltonian 
(Eq. (\ref{BRGHam})) the operators ${\cal O}_9$ and ${\cal O}_{10}$ have to be included. They dominate the decay rate in the region of 
higher dilepton mass $s_l$. We use in our calculation the Wilson coefficients $c_7 = -0.315$, $c_9 = 4.224$ and $c_{10} = -4.642$.

The resulting differential decay rate is written in a compact form in Eq. (3.7) of \cite{KSSS}. The effects of the long-distance contributions
to this decay are incorporated by modifying the form factors $g_+(s_l)$ and $g_-(s_l)$ ($q^2 \equiv s_l$) in Eqs. (2.16) and (2.18) of \cite{KSSS}:
\footnote{For the form factors, we have used the parametrization in Table IV of \cite{KSSS}. However, the normalization has been
updated to take account of the value  $g_+(0)|_{B^- \to \rho^-} = - T_1^{B^- \to \rho^-}(0) = -0.27$ \cite{Beyer01} instead of the value $-0.18$ used in 
\cite{KSSS}. Note that $g_\pm(0) \equiv g_\pm(0)|_{B^0 \to \rho^0} = \frac{1}{\sqrt{2}} g_\pm(0)|_{B^- \to \rho^-}$.}
\begin{eqnarray}
g_+(s_l) & \to & g_+(s_l) - \xi g_{\rm LD}(s_l),\\
g_-(s_l) & \to & g_-(s_l) + \xi g_{\rm LD}(s_l),
\end{eqnarray}
where
\begin{equation}
g_{\rm LD}(0) = \frac{V_{cb} V^{\ast}_{cd}}{V_{tb} V^{\ast}_{td}} \frac{f_{\rho^0 D^+ D^-} g_{B^0 D^+ D^-} }{4 c_7 m^2_{D^+} (m_b - m_d)}
F(m^2_{B^0}, m^2_{\rho^0}),
\end{equation}
and
\begin{equation}
\frac{g_{\rm LD}(s_l)}{g_{\rm LD}(0)} = G(m^2_{B^0}, s_l) \, F^D_{\rm em}(s_l).
\end{equation}
The function $G(m^2_{B^0}, s_l)$ describes the effects due to the triangle graph in Fig. (\ref{LDtriangle}) (assuming, for simplicity,
scalar external particles):
\begin{eqnarray}
G(m^2_{B^0}, s_l) & = & \left[ 1 - \frac{f_c \, \theta (c-1) + f_c^{\prime} \, \theta (1-c)}{f_a} \right] \left( 1-\frac{s_l}{m^2_{B^0}} \right)^{-1}, \\
f_c & = & - \left(\ln \left( \sqrt{c} + \sqrt{c-1} \right) - i \frac{\pi}{2} \right)^2, \nonumber \\
f_c^{\prime} & = & \arcsin^2 \left( \sqrt{c} \right), \nonumber \\
c & = & \frac{s_l}{4 m^2_{D^+}}. \nonumber
\end{eqnarray}
and $f_a$ as in Eq. (\ref{FB0Mrho}). The factor $F^D_{\rm em}(s_l)$ is the electromagnetic form factor of the $D$ meson in
vector dominance approximation:
\begin{eqnarray}
F^D_{\rm em}(s_l) & = & \frac{3}{2} \frac{1}{1 - \frac{s_l}{m^2_{\rho^0}} - i \frac{\Gamma_{\rho^0}}{m_{\rho^0}} \left(1 - \frac{4 m^2_{\pi}}{s_l}
\right)^{\frac{3}{2}} \theta (s_l - 4 m^2_{\pi} ) } \\
& & - \frac{1}{2} \frac{1}{1 - \frac{s_l}{m^2_{\omega}} - i \frac{\Gamma_{\omega}}{m_{\omega}} \theta (s_l - 9 m^2_{\pi} ) } \nonumber
\end{eqnarray}

\begin{figure}
\center
\psfrag{xi}[tl][tl][2]{$\xi$}
\psfrag{sl}[bl][bl][2]{$s_l / GeV^2$}
\resizebox{12cm}{!}{\rotatebox{-90}{\includegraphics{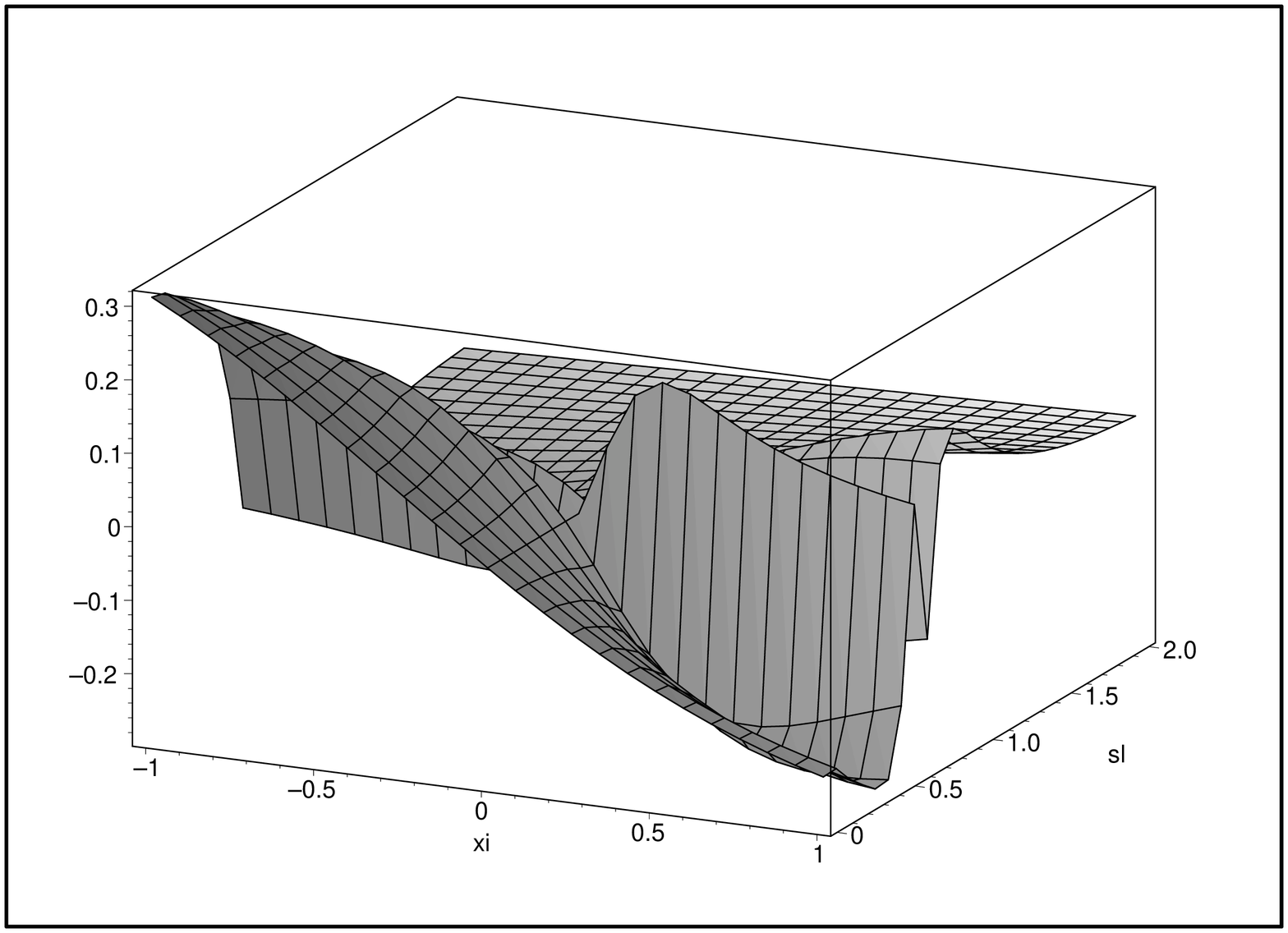}}}
\caption{The component $\Sigma_1$ as a function of the scale parameter $\xi$ of the long distance contribution 
and the dilepton energy $s_l$ in $\overline{B^0} \to \pi^+ \pi^- e^+ e^-$. \label{Sigma1Rho}}
\end{figure}

\begin{figure}
\center
\psfrag{xi}[tl][tl][2]{$\xi$}
\psfrag{sl}[bl][bl][2]{$s_l / GeV^2$}
\resizebox{12cm}{!}{\rotatebox{-90}{\includegraphics{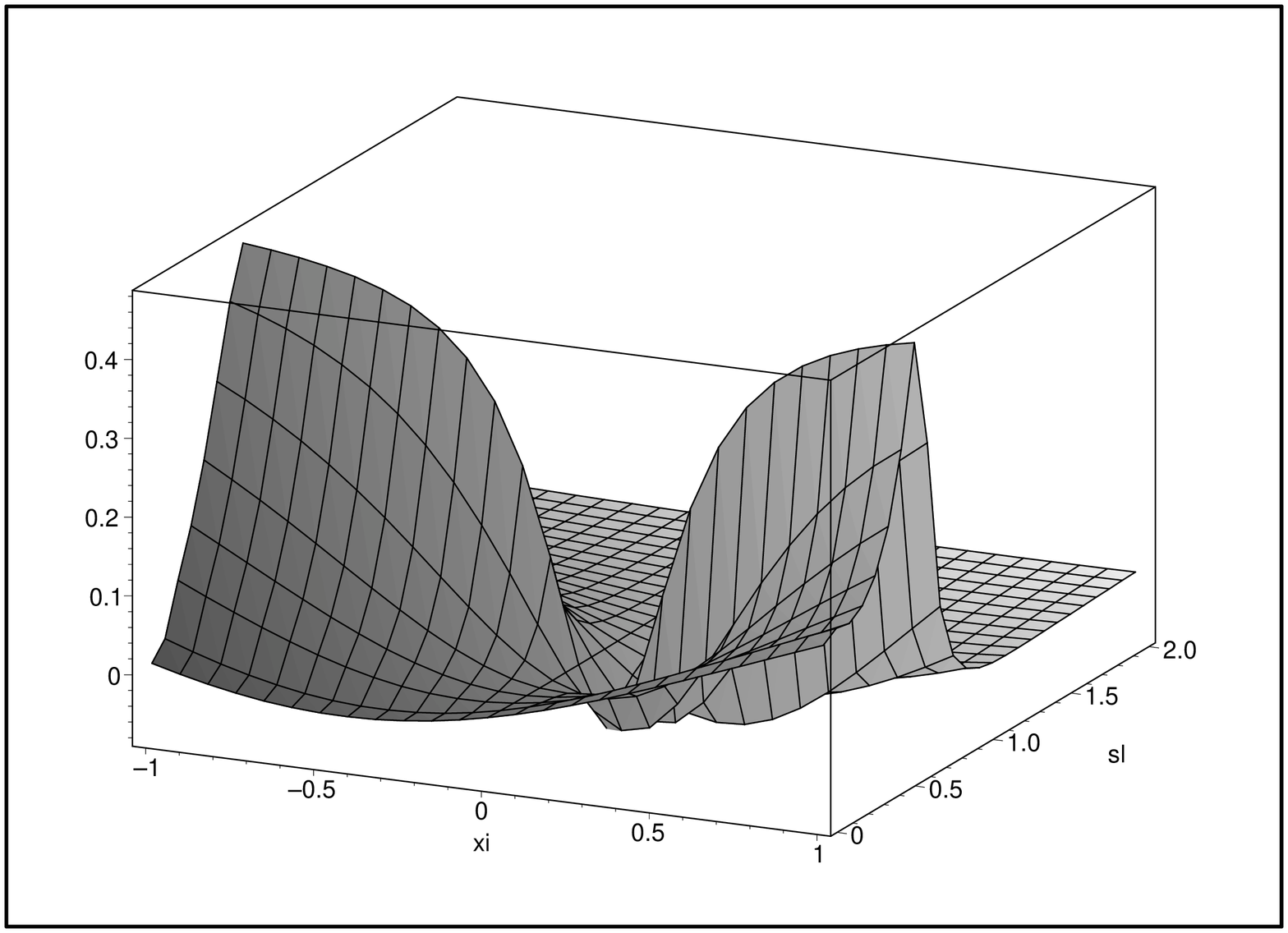}}}
\caption{The component $\Sigma_3$ as a function of the scale parameter $\xi$ of the long distance contribution 
and the dilepton energy $s_l$ in $\overline{B^0} \to \pi^+ \pi^- e^+ e^-$. \label{Sigma3Rho}}
\end{figure}

Integrating all variables but $\phi$ and $s_l$ in the differential decay rate yields Eq. (\ref{PhiDist}). The results for 
$\Sigma_1(s_l)$ and $\Sigma_3(s_l)$ are shown in Fig. (\ref{Sigma1Rho}) and Fig.(\ref{Sigma3Rho}), respectively. Comparing
the Stokes parameters $S_1$ to $\Sigma_1(0)$ and $S_3$ to $\Sigma_3(0)$ shows that they differ only by a factor of roughly 2. Both,
$\Sigma_1(s_l)$ and $\Sigma_3(s_l)$ can be dominated in the region $s_l < 2 \, GeV^2$ by the proposed long-distance effect,
depending on the choice of the parameter $\xi$. The branching ratio of $\overline{B^0} \to \rho^0 \gamma^{\ast} \to \pi^+ \pi^- e^+ e^-$
in the region $s_l < 2 \, GeV^2$ is found to be $(1.9, \, 1.3, \, 1.8) \times 10^{-8}$ for $\xi = (+\frac{1}{2}, \, 0, \, -\frac{1}{2})$. For $s_l > 2 \, GeV^2$ the
branching ratio is $2.7 \times 10^{-8}$, almost independent of $\xi$.

\section{Remarks on Stokes Parameter for $\overline{B^0} \to \overline{K^{\ast}} \gamma$}

Using a description for the short-distance amplitude in the decay $\overline{B^0} \to \overline{K^{\ast}} \gamma$ similar to that of
Eq. (\ref{AmplitudePengDom}), it immediately follows, as in pure short-distance $\overline{B^0} \to \rho^0 \gamma$, that
$A_{\rm CP} = 0$, $S_{1,3} = 0$, $S_2 = -1$. To embed the long-distance model in the decay $\overline{B^0} \to \overline{K^{\ast}} \gamma$
we used the same definitions as in $\overline{B^0} \to \rho^0 \gamma$ and made the following modifications, assuming $SU(3)$-symmetry:
The vertices $\overline{B^0} \to D^+ D^-$ and $\rho^0 D^+ D^-$ change to $\overline{B^0} \to D^+ D^-_s$ and $\overline{K^{\ast}} D^+ D^-_s$,
respectively. In the CKM matrix elements exchange $d \leftrightarrow s$ and align the form factors to the values found in the tables 
of $B \to \overline{K^{\ast}}$ decays \cite{FFPapers}.

The coupling $g_{B^0 D^+ D^-_s}$ can be calculated from the branching ratio $Br(\overline{B^0} \to D^+ D^-_s) = 9.6 \times 10^{-3}$
\cite{Neubert;Stech}, and $SU(3)$-symmetry gives $f_{\overline{K^{\ast}} D^+ D^-_s} = \sqrt{2} \, f_{\rho^0 D^+ D^-}$. 

\begin{figure}
\center
\psfrag{xi}[bl][bl][2]{$\xi$}
\resizebox{12cm}{!}{\rotatebox{-90}{\includegraphics{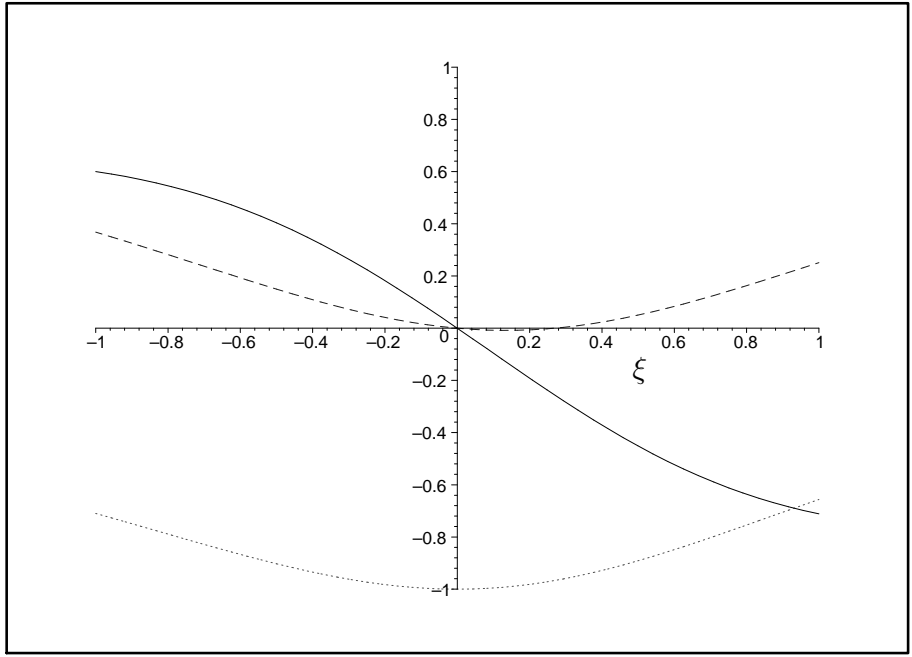}}}
\caption{The Stokesvector $\vec{S}$ as a function of the scale parameter $\xi$ of the long distance contribution 
in $\overline{B^0} \to K^{\ast} \gamma$. The solid line describes the $S_1$ component, the dotted line the $S_2$ 
component and the dashed line the $S_3$ component. \label{StokesPlotKG}}
\end{figure}

Since there is no relative weak phase (up to order $\lambda^4$) in $\overline{B^0} \to \overline{K^{\ast}} \gamma$ even after including the proposed
long-distance contribution, the  $CP$-asymmetry is $A_{\rm CP} = 0$ for all values of $\xi$. However, due to the absorptive part
of the triangle graph a strong (dynamical) phase is still present. In fact, it is essentially as large as in $\overline{B^0} \to \rho^0 \gamma$:
$\delta_{\rm dyn} \approx 97^{\circ}$. The presence of this phase can be seen in the Stokes vector components which are shown in
Fig. (\ref{StokesPlotKG}). For $\xi \neq 0$ the components $S_1$ and $S_3$ display a significant deviation from zero.

\begin{figure}
\center
\psfrag{xi}[tl][tl][2]{$\xi$}
\psfrag{sl}[bl][bl][2]{$s_l / GeV^2$}
\resizebox{12cm}{!}{\rotatebox{-90}{\includegraphics{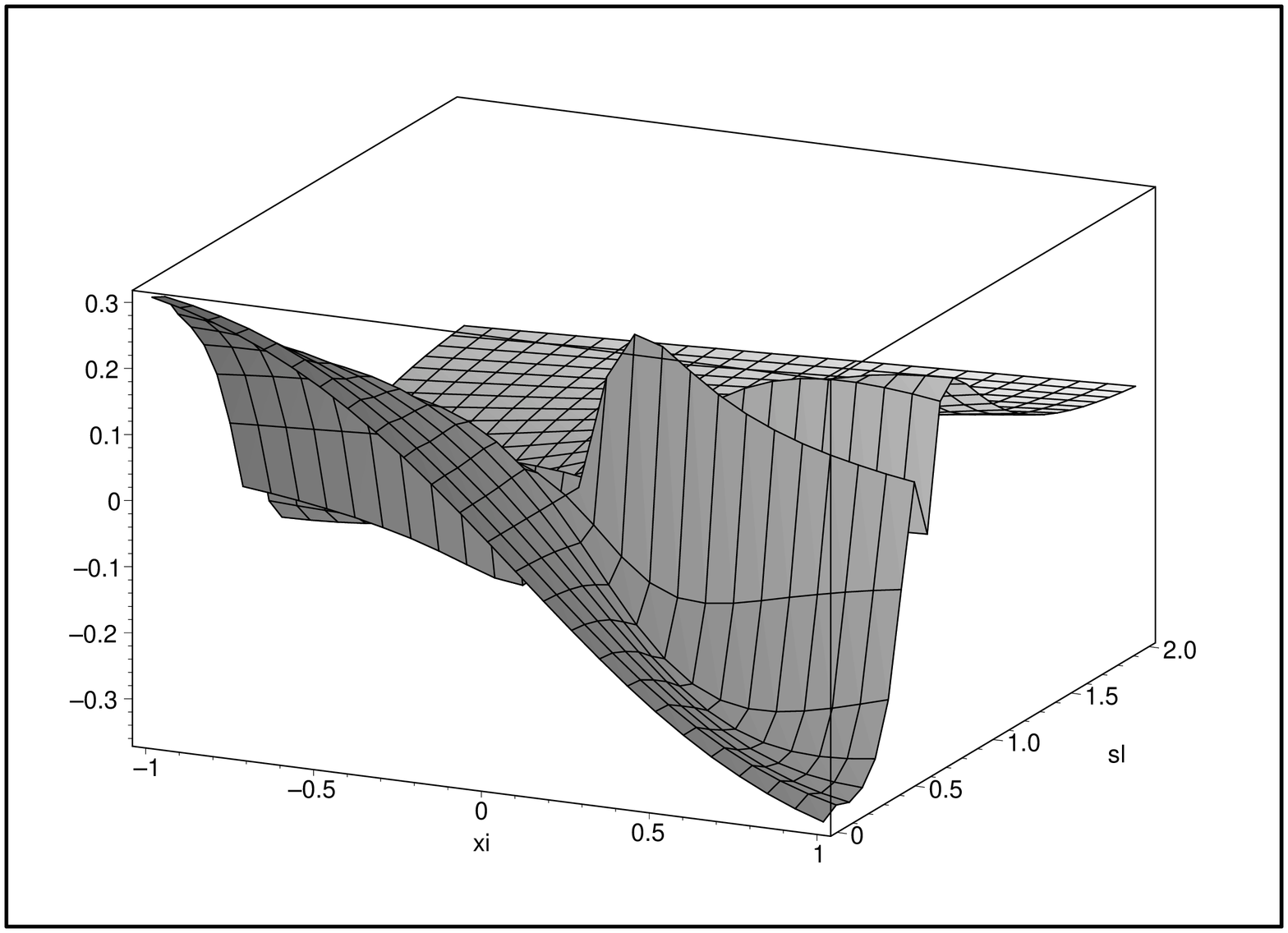}}}
\caption{The component $\Sigma_1$ as a function of the scale parameter $\xi$ of the long distance contribution 
and the dilepton energy $s_l$ in $\overline{B^0} \to \pi^+ K^- e^+ e^-$. \label{Sigma1K}}
\end{figure}

\begin{figure}
\center
\psfrag{xi}[tl][tl][2]{$\xi$}
\psfrag{sl}[bl][bl][2]{$s_l / GeV^2$}
\resizebox{12cm}{!}{\rotatebox{-90}{\includegraphics{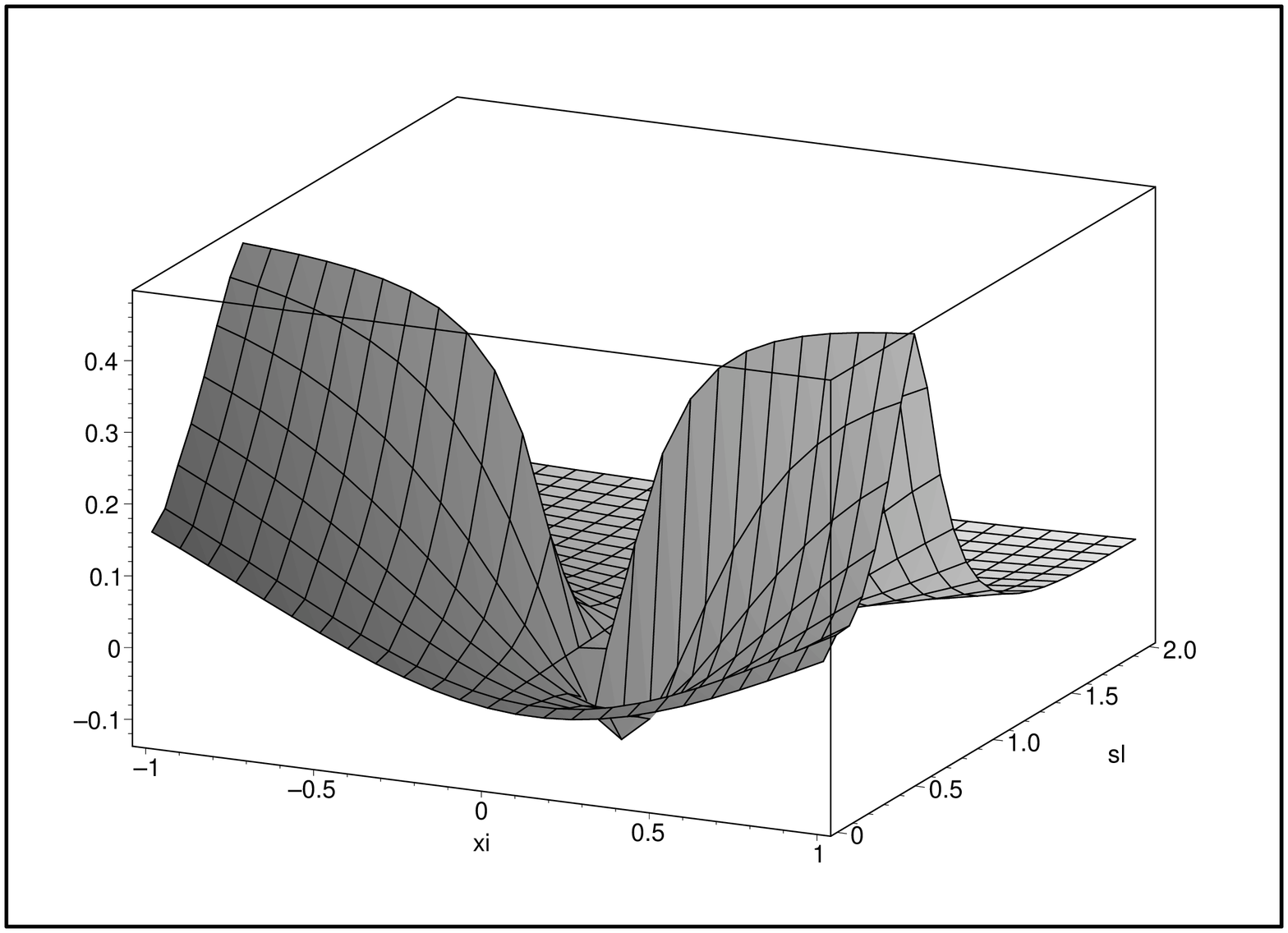}}}
\caption{The component $\Sigma_3$ as a function of the scale parameter $\xi$ of the long distance contribution 
and the dilepton energy $s_l$ in $\overline{B^0} \to \pi^+ K^- e^+ e^-$. \label{Sigma3K}}
\end{figure}

The Stokes vector components $S_1$ and $S_3$ can be detected in $\overline{B^0} \to \overline{K^{\ast}} \gamma^{\ast} \to \pi^+ K^- e^+ e^-$
\cite{KSSS} in the differential decay rate:
\begin{equation}
\frac{d \Gamma}{d s_l \, d \phi} \sim 1 - \left( \Sigma_1(s_l) \, \sin 2 \phi + \Sigma_3(s_l) \, \cos 2 \phi \right) ,
\end{equation}
which is derived in a way analogous to that in the decay $\overline{B^0} \to \pi^+ \pi^- e^+ e^-$ described before. The results
for $\Sigma_1$ and $\Sigma_3$ are shown in Figs. (\ref{Sigma1K}) and (\ref{Sigma3K}), respectively. Again, in the lower region of 
$s_l$ the long-distance contribution can play an important role depending on the parameter $\xi$. The branching ratio in the domain
$s_l < 2 \, GeV^2$ is $(0.8, \, 0.6, \, 0.9) \times 10^{-6}$ for $\xi = (+\frac{1}{2}, \, 0 , \, -\frac{1}{2})$, while for $s_l > 2 \, GeV^2$, the
corresponding value is $0.9 \times 10^{-6}$, essentially independent of $\xi$.

\section{Summary}

We have examined a long-distance contribution to the decay $\overline{B^0} \to \rho^0 \gamma$, which induces a non-zero $CP$-violating
asymmetry, $A_{\rm CP} \neq 0$. At the same time, the Stokes parameters $S_1$, $S_3$ of the photon acquire non-zero values that can 
be detected in the correlation of the $\pi^+ \pi^-$ and $e^+ e^-$ planes in the decay 
$\overline{B^0} \to \rho^0 \gamma^{\ast} \to \pi^+ \pi^- e^+ e^-$. The same long-distance mechanism has been examined in the case of
$\overline{B^0} \to \overline{K^{\ast}} \gamma$. Although $A_{\rm CP}$ remains zero in this case, significant effects due to the Stokes
parameters $S_1$, $S_3$ are predicted in the correlation of the hadron and lepton planes in the Dalitz pair process 
$\overline{B^0} \to \overline{K^{\ast}} \gamma^{\ast} \to \pi^+ K^- e^+ e^-$.


\begin{thebibliography}{99}
\bibitem{Beyer01} M.~Beyer, D.~Melikhov, N.~Nikitin and B.~Stech, Phys. Rev. D {\bf 64}, 094006 (2001).
\bibitem{Sehgal;Leusen} L.~M.~Sehgal and J.~van Leusen, Phys. Rev. Lett. {\bf 83}, 4933 (1999).
\bibitem{Grinstein;Ali} B.~Grinstein, D.~Pirjol, Phys. Rev. D {\bf 62}, 093002 (2000); A.~Ali, A.~Y.~Parkhomenko Eur. Phys. J. C {\bf 23}, 89 (2002).
\bibitem{CELZZ} D.~Choudhury and J.~Ellis, Phys. Lett. B {\bf 433}, 102 (1998); W.~Liu, B.~Zhang and H.~Zheng, Phys. Lett. B {\bf 461}, 295 (1999).
\bibitem{CharmPeng} M.~Ciuchini, R.~Contino, E.~Franco, G.~Martinelli and L.~Silvestrini, Nucl. Phys. B {\bf 512} 3 (1998); {\it ibid.} B {\bf 531} 656 (1998);
A.~J.~Buras and L.~Silvestrini, Nucl. Phys. B {\bf 569} 3 (2000).
\bibitem{FFPapers} M.~Wirbel, B.~Stech and C.~Bauer, Z. Phys. C {\bf 29}, 637 (1985); D.~Melikhov, Phys. Rev. D {\bf 53},
2460 (1996); D {\bf 56}, 7089 (1997); M.~Beyer, D.~Melikhov, Phys. Lett. B {\bf 452}, 121 (1999); D.~Melikhov, B.~Stech,
Phys. Rev. D {\bf 62}, 014066 (2000).
\bibitem{FSmith} J.~Pestieau, C.~Smith and S.~Trine, Int. J. Mod. Phys. A {\bf 17}, 1355 (2002);
C.~Smith, Ph. D. dissertation, Appendix A, Universit\a'e Catholique de Louvain, Louvain-la-Neuve, May 2002;
see also L.~M.~Sehgal, Phys. Rev. D {\bf 7}, 3303 (1973).
\bibitem{Browder} T.~Browder, talk at {\it Lepton \& Photon 2003}, Batavia, Illinois, 2003.
\bibitem{BaBarColl} Babar Collaboration, B.~Aubert {\it et al.}, Phys. Rev. Lett. {\bf 91}, 131801 (2003).
\bibitem{KSSS} F.~Kr\"uger, L.~M.~Sehgal, N.~Sinha and R.~Sinha, Phys. Rev. D {\bf 61}, 114028 (2000); {\it ibid.}
{\bf 63}, 019901(E) (2000).
\bibitem{Neubert;Stech} M.~Neubert and B.~Stech, in ``Heavy Flavours", $2^{\underline{{\rm nd}}}$ Edition, edited
by A.~J.~Buras and M.~Lindner (World Scientific, Singapore); hep-ph/9705292.
\end{thebibliography}
\end{document}